\documentclass[a4,11pt]{cp}  
\usepackage{graphics}
\usepackage{bm}% bold math
\usepackage{graphicx,axodraw}
\usepackage{amsmath,amsxtra,amssymb,latexsym, amscd,color}
\usepackage[mathscr]{eucal}
\def\vsm{\vskip0.1cm}
\def\titles#1{\title{\large\bf #1}}
\def\authord#1#2{\Mr{#1}\\
\textit{#2}\vsm}

\def\Mr{\MakeUppercase}

\def\cc{$^{12}$C+$^{12}$C}
\def\co{$^{12}$C+$^{16}$O}
\def\oo{$^{16}$O+$^{16}$O}
\def\AA{nucleus-nucleus\ }
\begin{document}
%PAGE_NUMBER%%%%%%%%%%%%%%%%%%%%%%%%%%%%%%%%%%%%%%%%%%%%%%%%%%%%%%%%
\Year{2015} \Page{1}\Endpage{10}
\titles{MEAN-FIELD STUDY OF $^{12}$C+$^{12}$C FUSION}
%AUTHORS AND ADDRESSES %%%%%%%%%%%%%%%%%%%%%%%%%%%%%%%%%%%%%%%%%%%%%%
\author{%--------------------------------------------------------------------------------------------------------------------------------
\authord{LE HOANG CHIEN}{Department of Nuclear Physics, Faculty of Physics and
Engineering Physics, University of Natural Science, Ho Chi Minh City, Vietnam}
\authord{DO CONG CUONG, DAO TIEN KHOA}{Institute for Nuclear Science and
Technology, VINATOM, Hanoi, Vietnam}}%-----------------------------------------------------------------------------------------------------------------------------------------
\date{today}
\maketitle
%SHORTEN NAME OF AUTHORS AND TITLE OF THE ARTICLE%%%%%%%%%%%%%%%%%%%%%%%%%%%%%%%%
\markboth{L.~H. CHIEN, D.~C. CUONG, D.~T. KHOA}{A MEAN-FIELD STUDY OF
THE $^{12}$C + $^{12}$C FUSION}

\begin{abstract}
The nuclear mean-field potential arising from the \cc\ interaction at the low 
energies relevant for the astrophysical carbon burning process has been  
constructed within the double-folding model, using the realistic nuclear
ground-state density of the $^{12}$C nucleus and the effective M3Y nucleon-nucleon
(NN) interaction constructed from the G-matrix of the Paris (free) NN potential. 
To explore the nuclear medium effect, both the original density independent 
M3Y-Paris interaction and its density dependent CDM3Y6 version have been 
used in the folding model calculation of the \cc\ potential. The folded potentials 
at the different energies were used in the optical model description of the elastic 
\cc\ scattering at the energies around and below the Coulomb barrier, as well as 
in the barrier penetration model to estimate the fusion cross section and 
astrophysical $S$ factor of the \cc\ reactions at the low energies. The obtained 
results are in good agreement with experimental data over a wide range of
energies.

\end{abstract}
 
\section{\Mr{Introduction}}
\label{intro}

The \cc\ fusion plays an important role in the whole chain of nucleosynthesis 
processes during stellar evolution, as the main nuclear reaction governing 
the carbon burning process in the young massive stars that generates the
heavier elements or the pycnonuclear reaction that leads a carbon-oxygen white 
dwarf to the type Ia supernova explosion \cite{Fowler84,Rolfs88, Iliadis15}. 
A general scenario for a massive star of about ten times the solar mass implies 
that after the helium burning process its core consists predominantly 
of the $^{12}$C and $^{16}$O ashes. As soon as this core begins to collapse 
gravitationally igniting the $^{12}$C and $^{16}$O ashes into the \cc, \co, and 
\oo\ fusion reactions, the first reaction is more favorable because it has the 
lowest Coulomb barrier. It also generates the heavier nuclei like $^{23}$Na, 
$^{20}$Ne, and $^{23}$Mg for the next burning stage of the stellar evolution.  
In fact, at the typical temperatures and densities in the outer-shell region 
of $T\approx 10^9$ K and $\rho\approx 10^5$ g/cm$^{3}$ respectively, the \cc\ 
fusion forms the $^{24}$Mg compound nucleus with mass difference between the \cc\
system and $^{24}$Mg nucleus of about 14 MeV. Therefore, the compound $^{24}$Mg$^*$ 
nucleus is highly excited and has a large number of overlapping states with 
the partial widths of the light particle emissions (neutron, proton and $\alpha$) 
dominating the $\gamma$-ray emission width. The main decay products of the
compound $^{24}$Mg$^*$ nucleus are $^{23}$Na, $^{20}$Ne, and $^{23}$Mg in the 
$^{12}$C($^{12}$C,$p)^{23}$Na (Q = 2241 keV), 
$^{12}$C($^{12}$C,$\alpha)^{20}$Ne (Q = 4617 keV), and
$^{12}$C($^{12}$C,$n)^{23}$Mg (Q = - 2599 keV) reaction channels, respectively, 
while the remaining processes such as $^{12}$C($^{12}$C,$\gamma)^{24}$Mg,
$^{12}$C($^{12}$C,$^{8}$Be)$^{16}$O are less important at astrophysical
energies \cite{Iliadis15}. In these channels, the $^{12}$C($^{12}$C,$p)^{23}$Na
and $^{12}$C($^{12}$C,$\alpha)^{20}$Ne reactions dominate the total \cc\
fusion cross section with about equal probabilities for proton and
$\alpha$ emissions.

In astrophysical conditions, the effective thermal energy of $^{12}$C
is approximately 2 MeV \cite{Fowler84} while the Coulomb barrier of the
\cc\ system is around 8 MeV which substantially lowers the probability of 
\cc\ fusion in such conditions. A narrow window for \cc\ fusion becomes 
possible thanks to the quantum tunneling effect that allows the two $^{12}$C 
nuclei to penetrate the Coulomb barrier without the need of having sufficient 
energy to overcome it \cite{Bribbin00}. The \cc\ fusion caused by the tunnel 
effect has been reasonably described by the barrier penetration model (BPM) 
\cite{Zettili09, Wong73, Vaz-Alex-Sat81}, which is used by many authors to
estimate nuclear reaction rates in stars. Typically, the nuclear reaction 
rate, a vital input for the study of stellar evolution, is expressed in 
terms of the astrophysical $S$ factor \cite{Fowler75}
\begin{equation}
 S = E_\textrm{c.m.}~\sigma_\textrm{R}~\textrm{exp} (2 \pi \eta),
 \label{e0}
\end{equation}
here $E_\textrm{c.m.}$ is the center-of-mass (c.m.) kinetic energy (in MeV) in
the entrance channel, $\sigma_\textrm{R}$ is the total reaction cross section 
(in barn) and $\eta$ is the Sommerfeld parameter determined as
\begin{equation}
 2 \pi \eta = 2 \pi \frac{Z_1 Z_2 e^2}{\hbar \nu} =
\frac{87.2}{\sqrt{E_\textrm{c.m.}\textrm{(MeV)}}},
\end{equation}
where $\nu$ is the relative velocity of the colliding nuclei. The astrophysical
$S$ factor is an important quantity introduced to describe 
the rate of a specific reaction in nuclear astrophysics studies \cite{Kunz96}. 
At very low energies, typical for nuclear astrophysics processes, the cross 
sections (or the astrophysical $S$ factors) of the charged-induced reactions are 
extremely difficult to measure in the laboratory because of the the repulsive Coulomb 
barrier that reduces the $S$ factor substantially. Therefore, it is important to have 
a reliable theoretical model to evaluate the astrophysical $S$ factors of
different nuclear reactions in the stellar energy region.

Because the \cc\ fusion reaction is an important part of the star evolution 
and still not fully understood, it has motivated many studies during the last 
four decades \cite{Patterson69, Mazarakis73, High77, Kettner80, Treu80, Becker81, 
Dasma82, Aguilera06}. The cross section of the \cc\ fusion reaction was calculated 
by different authors within the BPM framework using the different models of
the \cc\ potential \cite{Kondo78, Gasques04, Gasques05, Notani12,Aziz15}. However, 
the physics origin of the rapidly fluctuating \cc\ fusion cross section observed
at the lowest energies remains unexplained and needs to be further investigated.

In general, the \AA potential in the low-energy region can be naturally
associated with the nuclear mean field formed during the dinuclear collision
\cite{Bran-Sat97}. As a result, the so-called double folding model (DFM) 
which evaluates the \AA potential as the Hartree-Fock potential uses 
a realistic effective nucleon-nucleon (NN) interaction and the nuclear density
distributions of the two colliding nuclei \cite{Sat-Love79,Khoa00}. In the present 
paper, we explore the applicability of the DFM to determine the nuclear mean-field 
potential of the \cc\ system in the very low energy range (2-10 MeV), typical
of \cc\ fusion, using both the original M3Y-Paris interaction
\cite{Anan83} (constructed to reproduce the G-matrix elements of the Paris NN
potential \cite{Lacombe80} in an oscillator basis) and its CDM3Y6 density
dependent version \cite{Khoa97}. The \cc\ potential obtained in the DFM is
further used in the BPM to calculate the cross section and astrophysical $S$
factor of the \cc\ fusion reaction. 

The paper is organized as follows. In the next section, we give a brief
introduction to the theoretical methods used in this paper. The numerical
results and discussions are given in the Sec.~III. The last section 
summarizes the main results of the present work.

\section{\Mr{Theoretical methods}}

\subsection{The WKB method in the barrier penetration model}
\label{wkb}
The Wentzel-Kramers-Brillouin (WKB) method is well known to provide a
semi-classical approximation for the solution of the Schr\"odinger 
equation. As such, the WKB method has been used to elaborate the physics 
treatment of the BPM for \AA interacting systems at very low 
energies, where the nuclear mean-field potentials vary slowly over a spatial 
region of the order of the system wavelength \cite{Zettili09, Gasques04,
Gasques05}. In particular, the \cc\ fusion reaction can be studied within the 
BPM based on the simple treatment of the WKB method.

In general, the \AA interaction potential consists of the nuclear, centrifugal, 
and Coulomb terms
\begin{equation}
 V(r) = V_\textrm{N}(r)+\frac{l(l+1)\hbar^2}{2 \mu r^2}+V_\textrm{C}(r),
 \label{e1}
\end{equation}
where $l$ is the orbital angular momentum and $\mu=\dfrac{mA}{2}$ 
is the reduced mass of the \AA system, and $m$ is the free nucleon mass $m$. 
The Coulomb potential $V_\textrm{C}$ is usually assumed \cite{Sat83} as
\begin{eqnarray}
 V_\textrm{C}(r) = \begin{cases}
 \dfrac{Z^2e^2}{r} & \mbox{if}\ r > R_\textrm{c} \\
 \left(3-\dfrac{r^2}{{R^2_\textrm{c}}}\right)\dfrac{Z^2 e^2}{2R_\textrm{c}}
&\mbox{if}\ r\leqslant R_\textrm{c}
\label{e2}
\end{cases}
\end{eqnarray}
where $R_\textrm{c}=2 r_c A^{1/3}$ with $A$ being the mass 
number, and $r_c=0.95$ fm. 
The $l$-dependent centrifugal potential is that arising in the Schr\"odinger 
equation with spherically symmetric central potential.  
The nuclear potential $V_\textrm{N}$ given by the DFM calculation is used 
in the present work to determine the total \AA potential.

Within the BPM \cite{Wong73}, the fusion cross section of the particle flux 
transmitted through the Coulomb barrier is obtained from the $l$-dependent 
transmission coefficients $T_l$ as
\begin{equation}
 \sigma_\textrm{R} = \dfrac{\pi}{k^2} {\sum_0^{l_\textrm{cr}}}(2l+1)T_l,
\label{e4}
\end{equation}
where $k$ is the relative momentum, $l_\textrm{cr}$ is the critical angular
momentum corresponding to the largest value of the orbital angular momentum 
that reproduces both the pocket and barrier of the total \AA potential 
(\ref{e1}). $V_{Bl}$ is the barrier height, i.e., the value of the total \AA potential 
at the barrier radius $V(r=R_{Bl})$, which is different from the Coulomb
barrier.

For the partial waves $l$ with $V_{Bl} < E_\textrm{c.m.}$, the shape of the \AA 
potential around $R_{Bl}$ can be approximated as a parabola with the curvature 
determined as
\begin{equation}
 \hbar\omega_l = \bigg | \dfrac{\hbar^2}{\mu} \dfrac{d^2 V}{dr^2}
\bigg |_{R_{Bl}}^{1/2}.
\label{e5}
\end{equation}
Then, the transmission coefficient $T_l$ is obtained from the Hill-Wheeler 
formula \cite{Hill-Well53} as
\begin{equation}
 T_l= \left[ 1 + \exp \left( {\frac{ 2\pi [V_{Bl} -
 E_\textrm{c.m.}] }{\hbar \omega_l}} \right) \right]^{-1}.
\label{e6}
\end{equation}
For  the partial waves $l$ with $V_{Bl} > E_\textrm{c.m.},\ T_l$ is determined based 
on the WKB approximation
\begin{equation}
 T_l = [1 + \exp(S_l)]^{-1},
\label{e7}
\end{equation}
\begin{equation}
S_l = \int^{r_2}_{r_1} \sqrt{\frac{8\mu}{\hbar^2}
\bigg[V(r)-E_\textrm{c.m.}\bigg]}dr,
\label{e8}
\end{equation}
here $r_1, r_2$ are the radii of the classical turning points where 
$V(r_{1(2)})=E_\textrm{c.m.}$.

\subsection{Double-folding model of the \AA potential}
\label{folding}
In the framework of the BPM and the nuclear optical model \cite{Bec69}, 
the nuclear part of the total \AA potential is the most important input. 
From the physics point of view, it is always of interest to determine 
$V_\textrm{N}$ starting from the nucleon degrees of freedom, and the double-folding
model \cite{Sat-Love79,Khoa00} is the most commonly used approach for that
purpose. 
In this model, $V_\textrm{N}$ is evaluated as the Hartree-Fock (HF) 
potential with an appropriately chosen effective NN interaction between nucleons
in the target and projetile
\begin{equation}
 V_{\rm N}=V_{\rm D}+V_{\rm EX}=\sum_{i\in A_1,j\in A_2}
[\langle ij|v_{\rm D}|ij\rangle+\langle ij|v_{\rm EX}|ji\rangle], \label{fd1}
\end{equation}
Treating explicitly the single-nucleon wave functions in the HF potential
(\ref{fd1}), the local direct term is reduced to a double-folding
integration of the densities of the two colliding nuclei with the direct
part of the NN interaction  
\begin{equation}
 V_{\rm D}(r)=\int\rho_1(\bm{r}_1)\rho_2(\bm{r}_2)
 v_{\rm D}(\rho,s)d^3r_1 d^3r_2,\ \bm{s}=\bm{r}_2-\bm{r}_1+\bm{r}. \label{fd2}
\end{equation}
The antisymmetrization gives rise to the exchange term in Eq.~(\ref{fd1}) which
is, in general, nonlocal. An accurate local equivalent exchange potential can be 
obtained \cite{Khoa00} using the local WKB approximation \cite{Sat83} for the 
change in relative motion induced by the exchange of the spatial coordinates of each 
interacting nucleon pair
\begin{equation}
V_{\rm EX}(r) =\int \rho_1(\bm{r}_1,\bm{r}_1+\bm{s})\rho_2(\bm{r}_2,\bm{r}_2-\bm{s})
v_{\rm EX}(\rho,s)\exp\left({i\bm{k}(r)\bm{s}}\over{M}\right)d^3r_1 d^3r_2.
\label{fd3}
\end{equation}
Here $\bm{k}(r)$ is determined as
\begin{equation}
 k^2(r)={{2\mu}\over{\hbar}^2}[E_{\rm c.m.}-V_{\rm N}(r)-V_{\rm C}(r)],
\label{fd4}
\end{equation}
where $M=2 A$, $V_{\rm N}(r)$ and $V_{\rm C}(r)$
are the nuclear and Coulomb parts of the total \AA potential, respectively,
and $\rho_{1(2)}(\bm{r},\bm{r}')$ is the single-nucleon density matrix.  
It can be seen from Eqs.~(\ref{fd1})-(\ref{fd4}) that the DFM calculation 
of the \AA potential (\ref{fd1}) is a self-consistent problem. Therefore, the 
calculation of $V_{\rm EX}$ is carried out iteratively based on a realistic
expansion method for the density matrix \cite{Khoa00}. 

Among different choices of the effective NN interaction, a density dependent
version of the M3Y-Paris interaction (dubbed as CDM3Y6 interaction \cite{Khoa97}) 
has been used quite successfully in the folding model analyses of elastic
and inelastic \AA scattering. The density dependent parameters of the CDM3Y6 
interaction were carefully adjusted in the HF scheme to reproduce the saturation
properties of nuclear matter \cite{Khoa97}. In the present work, both the CDM3Y6
and original density independent M3Y-Paris interactions were used in the DFM
calculation. To avoid a phenomenological choice of the imaginary part of the
nuclear optical potential, the CDM3Y6 interaction has been supplemented with a
realistic imaginary density dependence for the folding calculation of the
imaginary potential. The parameters of the imaginary density dependence have
been deduced at each energy based on the Brueckner Hartree-Fock results for the
nucleon optical potential in nuclear matter by Jeukenne, Lejeune and Mahaux,
widely known as the JLM potential \cite{Je77}. Given an accurate choice of the
effective NN interaction, the DFM can be applied successfully to calculate the
\AA potential only if the realistic nuclear densities were used in the folding
calculation (\ref{fd2})-(\ref{fd3}). In the present work, the two-parameter
Fermi function was used for the ground-state density of the $^{12}$C 
nucleus
\begin{equation}
 \rho(r)=\rho_0/\{1+\exp[(r-c)/a]\}.
\label{fd5}
\end{equation}
The parameters in Eq.~(\ref{fd5}) were chosen to reproduce reasonably the empirical 
nuclear root-mean-square radius based on elastic electron scattering data as
well as the radial shape of the nuclear density given by the shell model
calculations \cite{Sat-Love79,Khoa00}. Given the appropriate choice of the
ground-state density of $^{12}$C and realistic density dependent NN interaction,
the folded \cc\ potential (\ref{fd1})-(\ref{fd3}) represents the mean-field
potential \cite{Bran-Sat97} in the nuclear medium formed in the \cc\ collision.
As such, the folded \cc\ potential can be used as the nuclear optical potential
to study the elastic \cc\ scattering and to estimate the reaction rate of the
\cc\ fusion in the BPM.

\section{RESULTS AND DISCUSSIONS}
\label{Results}
\begin{figure}[bht]%[bht]
% \hspace{-3.0cm}%\vspace{1.5cm}
\includegraphics[angle=0,scale=0.9]{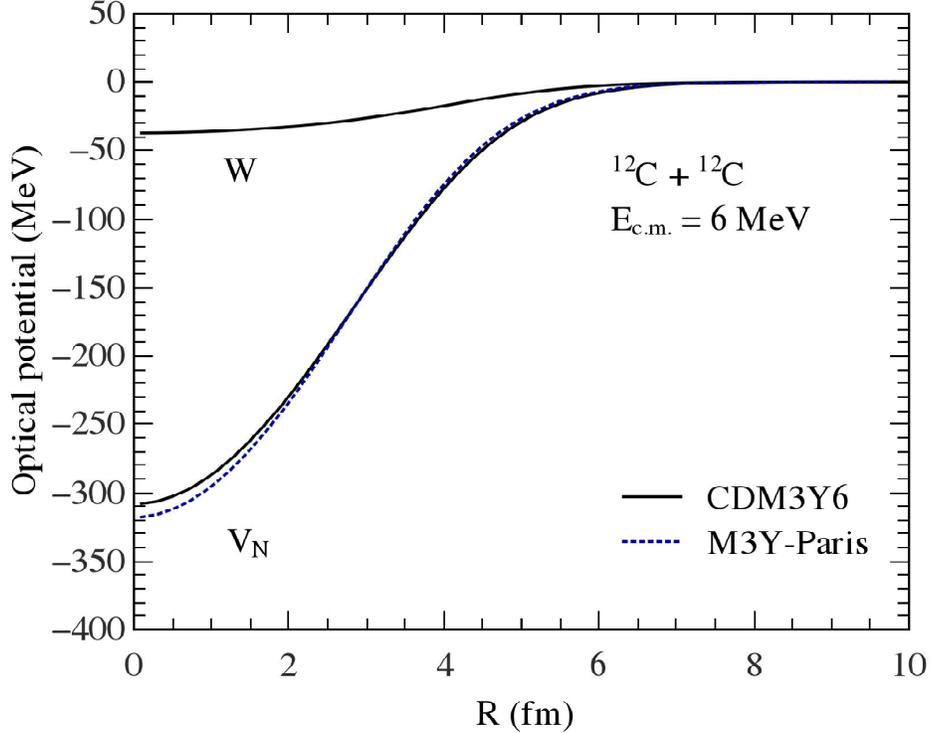} \vspace{-0.2cm}
\caption{The complex \cc\ optical potential folded with the density dependent CDM3Y6 
interaction (solid line) at $E_{\rm c.m.}=6$ MeV is compared with the real
potential folded with the density independent M3Y-Paris interaction (dashed
line).} 
 \label{f1}
\end{figure}

The reliability of the folded \cc\ potential should be tested first in the optical model 
description of elastic \cc\ scattering at low energies before using it in the
BPM to determine the astrophysical factor $S$ of \cc\ fusion. In the present
work we have analyzed the elastic \cc\ scattering data measured at energies 
$E_{\rm c.m.}=6-10$ MeV \cite{Treu80}, using the complex optical potential given 
by the DFM calculation (\ref{fd1})-(\ref{fd3}). The radial shapes of the real 
($V_{\rm N}$) and imaginary ($W$) potentials folded with the density dependent 
CDM3Y6 interaction, as shown in Fig.~\ref{f1}, are compared with the real
potential obtained with the density independent M3Y-Paris interaction. One can
see that the medium effects given by the density dependence of the NN
interaction make the real optical potential slightly shallower in the center but
more attractive at the potential surface.  

\begin{figure}[bh]%[bht]
% \hspace{-3.0cm}%\vspace{1.5cm}
\includegraphics[angle=0,scale=0.9]{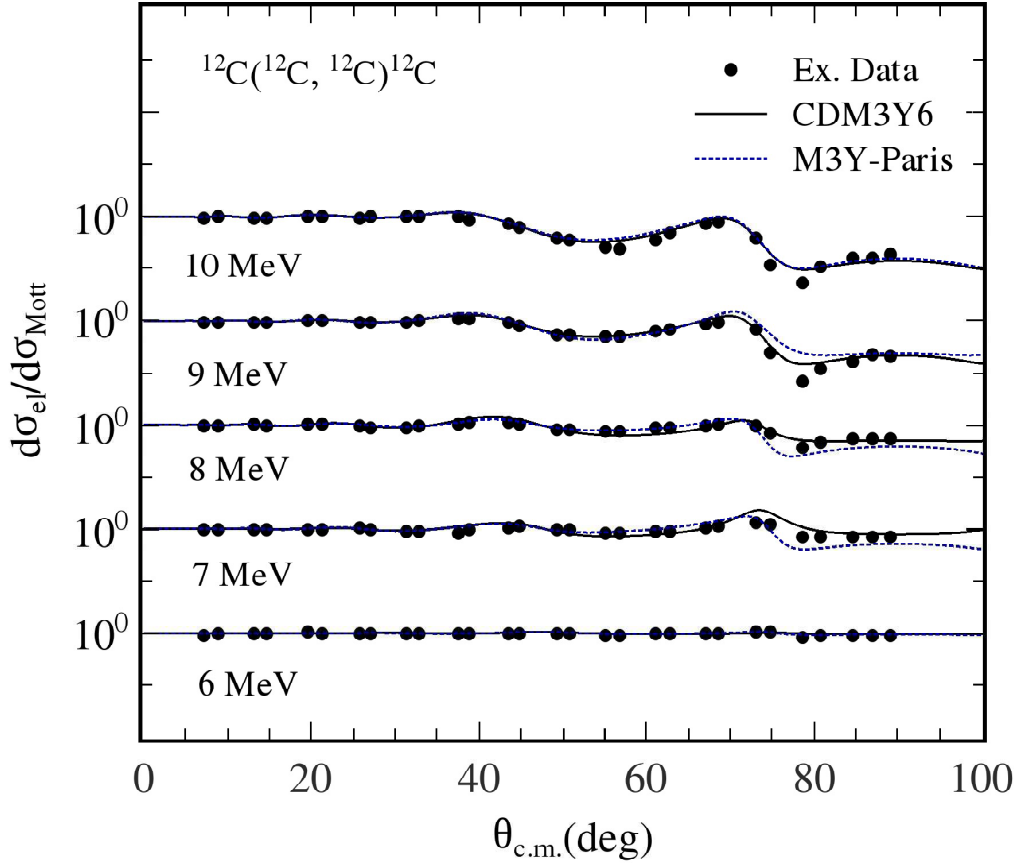} \vspace{-0.2cm}
\caption{The elastic \cc\ scattering cross sections (as ratio to the Mott 
cross section of the Coulomb scattering of the two identical charged
particles) at energies around the Coulomb barrier. The dashed and solid
curves are the results of the optical model calculation using the M3Y-Paris and
CDM3Y6 interactions, respectively. The experimental data are taken from
Ref.~\cite{Treu80}.} 
\label{f2}
\end{figure}

The (energy-dependent) complex folded CDM3Y6 potential can be used as the optical
potential to study elastic \cc\ scattering at low energies, relevant for
nuclear astrophysics. In the present work, we have considered five elastic 
scattering angular distributions measured in \cc\ collisions at energies around 
the Coulomb barrier \cite{Treu80}. To fine tune the complex strength of the optical
potential, a slight renormalization is usually adopted for the best optical model
fit of the experimental data. Thus, the complex optical potential used as 
input for the Schr\"odinger equation has the form
\begin{equation}
 U(r,E) = N_\textrm{r} V_{\rm N}(r,E) + iN_\textrm{i} W(r,E).
\end{equation}
Very good optical model description of the considered elastic \cc\ data has been
obtained with the complex folded CDM3Y6 potential renormalized by 
$N_\textrm{r}\approx 0.85$ and $N_\textrm{i}\approx 1.0$ (see Fig.~\ref{f2}). 
\begin{figure}[bht]
% \hspace{-3.0cm}%\vspace{1.5cm}
\includegraphics[angle=0,scale=0.85]{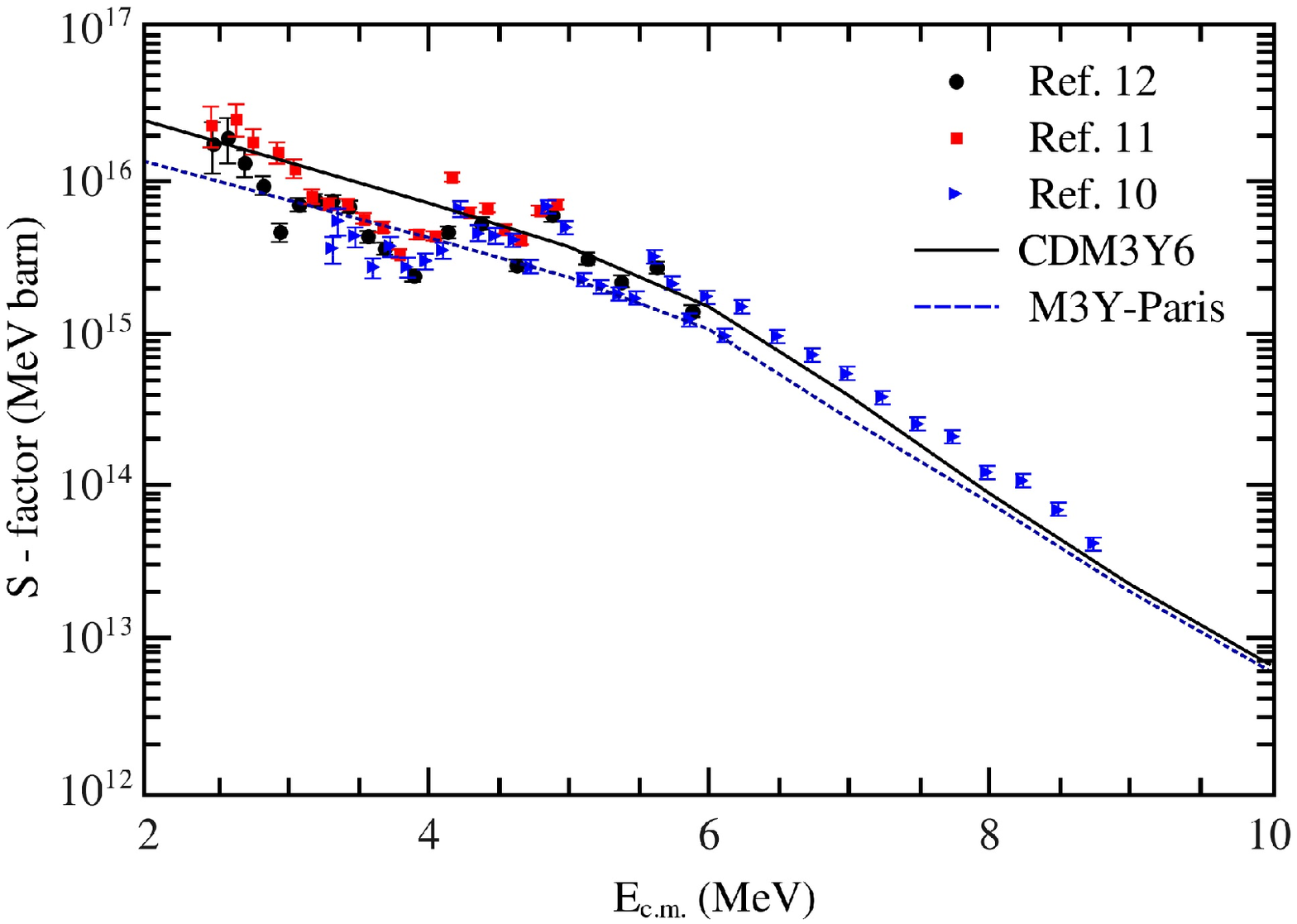} \vspace{0cm}
\caption{The astrophysical $S$ factor as a function of the center-of-mass energy. 
The dashed and solid curves represent the BPM results given by the M3Y-Paris 
and CDM3Y6 interactions, respectively. The experimental data are taken from 
Refs.~\cite{Patterson69,Mazarakis73, High77}.} \label{f3}
\end{figure}

One can see from the results plotted in Fig.~\ref{f2} that the complex folded 
CDM3Y6 potential gives a very good description of the elastic \cc\ data at
low energies. The effect on the real optical potential caused by the density 
dependence of the CDM3Y6 interaction shows up in the difference between 
the results given by the M3Y-Paris interaction (dashed line in Fig.~\ref{f2}) 
and those given by the CDM3Y6 interaction (solid line). Note that
these optical model calculations used the same imaginary part of the optical
potential obtained with the complex density dependent CDM3Y6 interaction. 
We found that the inclusion of the realistic density dependence into the 
effective NN interaction discussed widely in Refs.~\cite{Khoa00,Khoa97}
is also necessary for the good optical model description of elastic \cc\
scattering at low energies. It remains now to be seen whether this effect 
can also be observed in the BPM calculation of the \cc\ fusion reactions at very 
low energies.

The nuclear folded M3Y-Paris and CDM3Y6 potentials have been further used
in the BPM to evaluate the \cc\ fusion cross section using
Eqs.~(\ref{e1})-(\ref{e8}), and the results were used in Eq.~(\ref{e0}) to
calculate the astrophysical $S$ factor of the \cc\ fusion reaction. The results
obtained for the $S$ factor  are shown in Fig.~\ref{f3} and one can see a
reasonably good agreement of the BPM results with the experimental data
\cite{Patterson69, Mazarakis73, High77} over a wide range of energies. However,
the slight wiggling behavior of the experimental $S$ factor in the energy range
of 2 to 5 MeV (see Fig.~\ref{f3}) cannot be reproduced by the
BPM using the mean-field potential of the \cc\ system. Such an oscillation of
the $S$ 
factor in this energy range has been discussed as a resonant behavior of the \cc\
fusion reaction \cite{Kondo78, Almqvist60} caused by the relatively large
spacings and narrow widths of the $^{24}$Mg$^*$ compound levels \cite{Jiang13}. 
Although our mean-field approach to the \cc\ potential does not include any resonance
effect, the average description of the $S$ factor by the folded CDM3Y6 potential is 
quite satisfactory, so that the mean-field prediction of the \cc\ reaction rate is accurate
over the entire Gamow range.

The results of the BPM calculation shown in Fig.~\ref{f3} also show that the $S$
factors obtained with the density independent M3Y-Paris interaction are somewhat
lower than the experimental data and those obtained with the density dependent
CDM3Y6 interaction. Technically it is explained by the fact that the M3Y-Paris
potential provides a higher barrier in comparison with that given by the CDM3Y6
potential. Thus, the medium effects caused by the density dependence of the
effective NN interaction are not negligible in the BPM calculation of the $S$
factor, and this conclusion is natural in view of the carbon burning process
occurring in the dense baryon matter of very massive stars. 

\section{SUMMARY}
The nuclear mean-field potential arising in the \cc\ collision at very low energies
has been constructed in the double-folding model using the realistic nuclear
density and the complex density dependent CDM3Y6 interaction, based on the
original M3Y-Paris interaction. The complex folded \cc\ potential was used 
in the optical model to successfully describe the elastic \cc\ scattering at 
the low energies around the Coulomb barrier. This same potential was shown to
give also a realistic description of the astrophysical $S$ factor 
for the \cc\ fusion reaction over a wide range of the energies. 

The mean-field description of both the elastic scattering angular distributions 
and the $S$ factor of the \cc\ reaction at the low energies has shown a rather 
strong medium effect caused by the density dependence of the effective
NN interaction. The results obtained in the present work also confirmed 
the reliability of the double-folding model in the calculation of the total nuclear 
potential for the study of  the \cc\ fusion reaction in the low-energy region of the
nuclear astrophysical interest. 

The further use of the DFM in the calculation of both the optical potential and
inelastic scattering form factor is planned to be used within the framework of
the coupled channel formalism for the study of inelastic scattering with the 
final state of $^{12}$C nucleus in excited states, such as $2^+$ state at 4.44
MeV, $0^+_2$ state at 7.65 MeV, and $3^-$ state at 9.64 MeV. Besides, at the
low energies around the Coulomb barrier the dominant final states of the \cc\
reaction are $^{20}$Ne + $\alpha$, $^{23}$Na + $p$ and $^{16}$O + $^8$Be, and it
is of high interest to estimate their explicit contributions to the total \cc\
reaction cross section in this energy range.

\section*{ACKNOWLEDGMENT }
The present research has been supported, in part, by the National Foundation for 
Scientific and Technological Development (NAFOSTED Project No. 103.04-2014.76).


\begin{thebibliography}{99} %Phan trich-dan 
\bibitem{Fowler84} W. A. Fowler, {\em Rev. Mod. Phys.}, {\bf 56} (1984) 149.
\bibitem{Rolfs88} C. E. Rolfs, {\em Cauldrons in the Cosmos}, University of
Chicago Press, Chicago (1988).
\bibitem{Iliadis15} C. Iliadis, {\em Nuclear Physics of Stars}, Wiley-VCH
Press, Weinheim (2015).
\bibitem{Bribbin00} J. Gribbin, M. Gribbin, {\em Stardust}, Allen Lane The
Penguin Press, London (2000).
\bibitem{Zettili09} N. Zettili, {\em Quantum mechanics concepts and
applications}, A John Wiley and Son, University Press, London (2009).
\bibitem{Wong73} C. Y. Wong, {\em Phys. Rev. Lett.}, {\bf 31} (1973) 776.
\bibitem{Vaz-Alex-Sat81} L. C. Vaz, J. M. Alexander, and G. R. Satchler,
{\em Phys. Rep.}, {\bf 69} (1981) 373.
\bibitem{Fowler75} W. A. Fowler, G. R. Caughlin, and B. A. Zimmerman, {\em Annu.
Rev. Astron. Astrophys.}, {\bf 13} (1975) 69.
\bibitem{Kunz96}R. Kunz, S. Barth, A. Denker, H. W. Drotleff, J. W. Hammer, H.
Knee, and A. Mayer, {\em Phys. Rev. C}, {\bf 53} (1996) 2486.
\bibitem{Patterson69} J. A. Patterson, H. Winkler, and C. S. Zaidins,
{\em Astrophys. J.}, {\bf 157} (1969) 367.
\bibitem{Mazarakis73} M. G. Mazarakis, W. E. Stephens, {\em Phys.
Rev. C}, {\bf 7} (1973) 1280.
\bibitem{High77} M. D. High, B. Cujec, {\em Nucl. Phys. A}, {\bf 282}
(1977) 181.
\bibitem{Kettner80} K. U. Kettner, H. Lorenz-Wirzba, and C. Rolfs, {\em Z. Phys.
A}, {\bf 298} (1980) 65.
\bibitem{Treu80} W. Treu, H. Frohlich, W. Galster, P. Duck, and H. Voit,
{\em Phys. Rev. C}, {\bf 22} (1980) 2462.
\bibitem{Becker81} H. W. Becker, K. U. Kettner, C. Rolfs, and H. P. Trautvetter,
{\em Z. Phys. A}, {\bf 303} (1981) 305.
\bibitem{Dasma82} B. Dasmahapatra, B. Cujec, and F. Lahlou, {\em Nucl.
Phys. A}, {\bf 384} (1982) 257. 
\bibitem{Aguilera06} E. F. Aguilera {\em et al}., {\em Phys. Rev. C}, {\bf 73}
(2006) 064601.
\bibitem{Kondo78} Y. Kondo, T. Matsuse, and Y. Abe, {\em Prog.
Theo. Phys.}, {\bf 59} (1978) 465.
\bibitem{Gasques04} L. R. Gasques {\em et al.}, {\em Phys. Rev. C}, {\bf 69}
(2004) 034603.
\bibitem{Gasques05} L. R. Gasques {\em et al.}, {\em Phys. Rev. C}, {\bf 72}
(2005) 025806.
\bibitem{Notani12} M. Notani {\em et al}., {\em Phys. Rev. C}, {\bf 85} (2012)
014607.
\bibitem{Aziz15} A. A. Aziz, N. Yusof, M. Z. Firihu, and H. A. Kassim, {\em
Phys. Rev. C}, {\bf 91} (2015) 015811.
\bibitem{Bran-Sat97} M. E. Brandan, G. R. Satchler, {\em Phys. Rep.}, {\bf
285} (1997) 143.
\bibitem{Sat-Love79} G. R. Satchler, W. G. Love, {\em Phys. Rep.}, {\bf 55}
(1979) 183.
\bibitem{Khoa00} D. T. Khoa, G. R. Satchler, {\em Nucl. Phys. A}, {\bf 668}
(2000) 3.
\bibitem{Anan83} N. Anantaraman, H. Toki, and G. F. Bertsch, {\em Nucl.
Phys. A}, {\bf 398} (1983) 269.
\bibitem{Lacombe80} M. Lacombe {\em et al.}, {\em Phys. Rev. C}, {\bf 21} (1980)
861.
\bibitem{Khoa97} D. T. Khoa, G. R. Satchler, and W. von Oertzen, 
 {\em Phys. Rev. C}, {\bf 56} (1997) 954.
\bibitem{Sat83} G. R. Satchler, {\em Direct Nuclear Reactions},
 Clarendon Press, Oxford (1983).
\bibitem{Hill-Well53} D. L. Hill, J. A. Wheeler, {\em Phys.
Rev.}, {\bf 89} (1953) 1102.
\bibitem{Bec69} F. D. Becchetti, Jr. and G. W. Greenlees, {\em Phys. Rev.}, {\bf
182} (1969) 1190.%optical potential
\bibitem{Je77} J. P. Jeukenne, A. Lejeune, and C. Mahaux,
 {\em Phys. Rev. C}, {\bf 16} (1977) 80.
\bibitem{Almqvist60} E. Almqvist, D. A. Bormley, and J. A. Kuehner, {\em Phys.
Rev. Lett.}, {\bf 4} (1960) 515.
\bibitem{Jiang13} C. L. Jiang, B. B. Back, H. Esbensen, R. V. F. Janssens,
K. E. Rehm, and R. J. Charity, {\em Phys. Rev. Lett.}, {\bf 110} (2013) 072701.

\end{thebibliography}
\end{document}